\documentclass[a4paper]{jpconf}
\usepackage{graphicx}
\usepackage{amssymb,amsmath}
\usepackage{algorithm}
\usepackage{algpseudocode}
\usepackage[%
	breaklinks=true,
	colorlinks=true,
	linkcolor=blue,
	urlcolor=blue,
	citecolor=blue
]{hyperref}

\begin{document}
\title{Average cluster size inside sediment left after droplet desiccation}

\author{P\,A~Zolotarev$^{1,2}$, K\,S~Kolegov$^{1,2,3}$}

\address{$^1$Astrakhan State University,  20A Tatishchev St., Astrakhan, 414056, Russia}

\address{$^2$Landau Institute for Theoretical Physics Russian Academy of Sciences, 1-A  Academician Semenov Ave., Chernogolovka, 142432, Russia}

\address{$^3$Volga State University of Water Transport, Caspian Institute of Maritime and River Transport, 6 Nikolskaya St., Astrakhan, 414000, Russia}

\ead{konstantin.kolegov@asu.edu.ru}

\begin{abstract}
In this work, we continue to study the formation of particle chains (clusters) inside the annular sediment during the drying of a colloidal droplet on a substrate. The average value of the cluster size was determined after processing experimental data from other authors. We performed a series of calculations and found the value of the model parameter allowed to get numerical results agreed with the experiment. Also, a modification of the previously proposed algorithm is analyzed here.
\end{abstract}

\section{Introduction}
Evaporation-induced self-assembly (EISA) methods are useful in applications such as inkjet printing, production of photonic crystals, application of transparent conductive coatings, development of biosensors, and others~\cite{Kolegov2020}. Modeling the formation of colloidal particle precipitation structures is important. It allows us to understand the main mechanisms of such processes, identify key parameters, and study how to influence the system to obtain the required patterns. For example, the model~\cite{ZhaoM2019} describes the self-assembly of particles on the free surface of a droplet. The capillary interaction of particles and their transport by surface flow resulting from evaporation is considered in this paper. The authors~\cite{ZhaoM2019} performed calculations for spherical and triangle droplets using the lattice Boltzmann method (LBM). Numerical results showed the dependence of the sediment on the curvature of the free surface. Clusters of particles formed at the liquid-vapor interface appear on the substrate after complete evaporation of the liquid. Another mechanism for the formation of particle clusters as a result of their capillary attraction is their self-assembly on the substrate where the thickness of the liquid layer is less than the particle size~\cite{Wouters2020}. The calculations were performed using the modification of LBM. The feature of the algorithm is the ability to simulate the dynamics of soft particles~\cite{Wouters2020}. The mass transfer of hard  particles in a drying film was studied in~\cite{Chun2020}. The LBM-based model takes into account the hydrodynamic interaction of particles and their diffusion. The authors~\cite{Chun2020} performed calculations for different values of the Peclet number associated with different evaporation rates. At high values of the Peclet number, a layer of particles is formed near the free surface of the film due to the rapid movement of the two-phase boundary as a result of intense evaporation. A uniform distribution of particles is observed at a small value of the Peclet number. It is explained by diffusion transfer. The periodic predominance of capillary flow or solutal Marangoni flow sometimes leads to the formation of concentric precipitation of colloidal particles~\cite{Seo2020}. The authors~\cite{Seo2020} used the lattice model and the Monte Carlo method to study the effect of the surfactant concentration on this process. The magnetic interaction of particles in a liquid and the formation of chains was studied in~\cite{Darras2017}. The model is based on the Soft Sphere Discrete Element Method. At each time step, Newton's equations were solved to predict the motion of particles. Calculations were performed for small viscosity values~\cite{Darras2017}. The model~\cite{Yang2020} takes into account the diffusion and sedimentation of particles. These particles have no volume, but they have mass. The model is based on the phase field method and the Monte Carlo method. Two shapes of the droplet surface were considered: a spherical segment and an asymmetric shape from the experiment~\cite{Deegan1997}. In both cases, a large number of deposited particles per unit area near the periphery is shown (the coffee stain effect). The model predicts a significant accumulation of particles in areas of low curvature for an asymmetric droplet shape. In the experiment~\cite{Deegan1997}, the particles accumulated mainly in areas of high curvature. The reason for this is the capillary flow, which is not taken into account in the model~\cite{Yang2020}. The formation of particle chains (clusters) inside the annular sediment during the evaporation of a colloidal droplet on a substrate was described in~\cite{Kolegov2019}. The model takes into account diffusion, advection, and capillary attraction of particles. Capillary interaction of particles occurs near the boundary (fixation radius $R_f$) where the thickness of the liquid layer is comparable to the particle size. This process was mimicked as follows.  In one time step, the particle was shifted to the nearest neighboring particle in the small neighborhood $R_n$ at the presence of the last one. Calculations were performed for only one parameter value $R_n$. In the current work, we consider the set of parameter  $R_n$ values  and determine the value which leads the average cluster size is similar to the experiment~\cite{Park2006}. Also, here we analyze the modification of the algorithm associated with multiple attempts to shift those moving particles whose location has not changed in a time step as a result of a collision with a neighboring particle.

\section{Methods}

\subsection{Problem statement and model description}
Here we consider an axisymmetric sessile droplet on a substrate in the mode of a pinned three-phase boundary. Colloidal particles are suspended in the liquid. They are subject to transfer by flow, diffusion, and capillary attraction. At the initial time, the contact angle $\theta_0=\pi/18$. The contact radius of the droplet with the substrate $R=$ 60~$\mu$m. The radius of spherical particles $r_p=$ 0.35~$\mu$m. As the liquid dries, these particles precipitate. The deposit is a monolayer of particles since the droplet is thin (its height $h\ll R$). Time of full droplet evaporation $t_\mathrm{max}=$ 10~s. A more detailed description of the problem statement is given in~\cite{Kolegov2019}.

The mathematical model~\cite{Kolegov2019} is semi-discrete. Each particle is considered separately, but the liquid is described with a continuum approach. It is assumed that the particle velocity is equal to the fluid flow velocity, which is calculated by the formula
$$
 \bar v_r (r,t)=
 \frac{R}{4 r/R (t_\mathrm{max}-t)}\left[  \frac{1}{\sqrt{1-(r/R)^2}} - \left( 1-(r/R)^2 \right) \right].
$$
It was derived from the mass conservation law (references and formula derivation are given in~\cite{Kolegov2019}). The model considers the radial flow averaged over the drop height. Here $r$ and $t$ is the radial coordinate of the particle center and the current time of the process, respectively.
Coordinate of the particle at the next time step as a result of displacement by the flow
\begin{equation*}
 r_{\tau+1} =
 \begin{cases}
  r_\tau + \bar v_r \delta t, \; r_{\tau+1}\eqslantless R_f \\
  R_f,\text{ otherwise,}
 \end{cases}
\end{equation*}
where $\delta t$ is the time step and $\tau$ is the time step number.

Capillary forces act on the particles in the region of the fixation boundary $R_f (t) = \sqrt{R^2 - 4 r_p R/ \theta (t)}$. The time dependence of the contact angle is written as
$\theta(t) = \theta_0 \left( 1 - t/ t_\mathrm{max} \right)$. We mimic the capillary attraction of particles as follows. If there are particles in the vicinity of the current particle that have precipitated or are also subject to capillary attraction, then it moves close to the nearest neighboring particle. By neighborhood, we mean a circle with a radius $R_n$. Some particles precipitate if $r > R_f+r_p$. In this case, the local thickness of the liquid layer is less than the particle size, $h < 2r_p$.

The diffusion motion of particles is modeled by the Monte Carlo method. A random offset angle $\alpha \in [-\pi; \pi)$ is generated at each time step $\tau$. The new position of the particle is calculated using the formulas $x_{\tau +1}= x_\tau + \cos (\alpha)\sqrt{2
D \delta t}$ and $y_{\tau +1}= y_\tau + \sin (\alpha)\sqrt{2
D \delta t}$, where $x$ and $y$ are the Cartesian coordinates of the particle center. The diffusion coefficient is calculated using the Einstein formula $D = kT/(6\pi \eta r_p)$, where $k$ is the Boltzmann constant. The values of the temperature $T$ and the viscosity $\eta$ of the liquid are taken for water under normal room conditions ($D\approx 6\cdot 10^{-13}$ m$^2$/s). The value of the time step $\delta t =$ 0.1~ms was selected based on a series of computational experiments in such a way that the Einstein relation for the mean square  displacement $\langle l^2 \rangle = 2D t_{max}$ was fulfilled. Here $$\langle l^2 \rangle =\frac{1}{N_p} \sum_{i=1}^{N_p}\left( \left( x_i(t_\mathrm{max}) - x_i(0) \right)^2 + \left( y_i(t_\mathrm{max}) - y_i(0) \right )^2 \right)$$
and $N_p$ is the number of particles.

\subsection{Algorithm}
\begin{figure}[H]
	\begin{minipage}{0.3\textwidth}
		\includegraphics[width=0.99\textwidth]{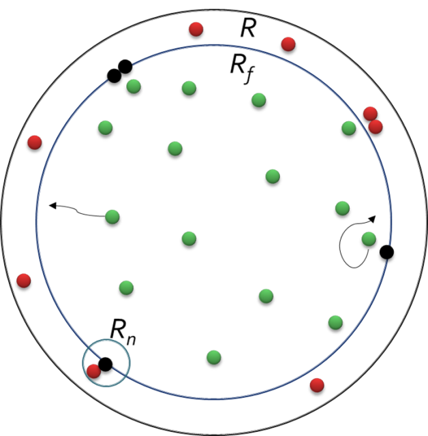}
	\end{minipage}\hspace{0.025\textwidth}
	\begin{minipage}{0.3\textwidth}
		\includegraphics[width=0.99\textwidth]{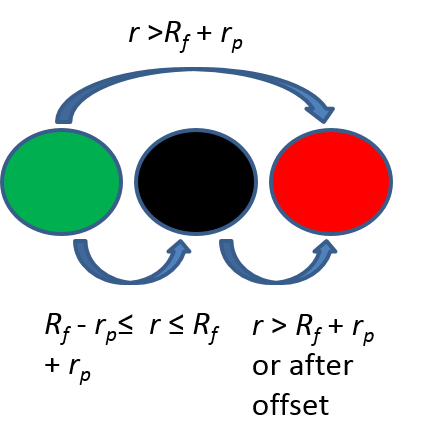}
	\end{minipage}
	\begin{minipage}{0.35\textwidth}
		\caption{\label{fig:algorithmDescription} Scheme to the algorithm description (left) and rules for changing the particle status (right).}
	\end{minipage}
\end{figure}

For each particle, we store coordinates and status in memory. Let's denote movable particles with green and black colors. Here green particles are subject to convective and diffusive transfer. We do not use a separate status for particles diffusing near $R_f$ to simplify the algorithm~\cite{Kolegov2019}. Capillary forces act on black particles. Unmovable particles that have precipitated are marked in red. The particle status changes according to the rules in Figure~\ref{fig:algorithmDescription} (right).

For more details about the algorithm, see Figure~\ref{fig:algorithmDescription} (left) and  pseudocode~\ref{alg:particleDynamicsModified}. At the beginning of the process, all particles are green by default. If a green particle touches the fixation radius, it is repainted black. The green particle cannot go beyond the fixation radius, unlike the black one. After all, the movement in that area is mainly due to capillary forces. But due to the movement of the fixation radius itself, the green particle may be behind it (after that, it will be repainted red). A black particle becomes red after it is beyond the moving boundary $R_f$ as a result of the motion of the last one or after it shifts due to capillary forces.
\begin{algorithm}[H]
	\caption{Particle dynamics algorithm.}
	\label{alg:particleDynamicsModified}
	\begin{algorithmic}[1]
		\State Problem parameters definition: $R_n$, $r_p$, $R$, $N_p$, and
		$t_\mathrm{max}$.
		\State Generation random coordinates of the particles
		$x_i$ and $y_i$ ($i \in [1;N_p]$).
		\State By default, all particles are marked green.
		\For {$\tau \leftarrow 1, t_\mathrm{max}/\delta t$}
		\State calculate $R_f$.
		\For {$i \leftarrow 1, N_p$}
		\State changing the particle status if necessary.
		\EndFor
		\While {(not all particles are displaced) $\And$ (there is a shift  of at least one particle)}
		\State shuffle the array of particle numbers.
		\For {$i \leftarrow 1, N_p$}
		\State red particles are skipped.
		\If {(green particle)}
		\State calculate the new particle coordinates due to diffusion.
		\If {(no collision)}
		\State move the particle.
		\EndIf
		\State Calculate the new particle coordinates due to advection.
		\If {(no collision)}
		\State move the particle.
		\EndIf
		\EndIf
		\If {(the particle is black) $\And$ (there is a black or red particle within $R_n$ )}
		\State Calculate the new particle coordinates due to the capillary shift.
		\If {(no collision)}
		\State move the particle close to the nearest one.
		\EndIf
		\EndIf
		
		\EndFor
		\EndWhile
		\State Write the particle coordinates and colors to a file for the current time step.
		\EndFor
	\end{algorithmic}
\end{algorithm}

At each time step, we calculate the new value of $R_f$. Then we go through all the non-red particles and check the conditions for changing their status. Next, we shuffle the numbers of particles so that their subsequent search is random. After that, we repeatedly iterate over the moving particles that have not yet moved at the current time step and try to move each of them in turn. Failure to shift can be caused by a collision (neighbor particle overlapping) or the exit of a green particle beyond the $R_f$ boundary as a result of a diffusive motion. Repeated iterations are completed if all the moving particles were displaced at the current time step or there were no displacements at all during a repeat of the loop. To check that there is no collision of a particle with any other, it is advisable to consider the distance only to the nearest neighbors. To do this, we have to store additionally and periodically update information about subdomains containing particles (numbers of particles included in the subdomain).

\subsection{Determination of cluster sizes}
To determine the value of the average cluster size, an experimental picture was taken from~\cite{Park2006} for further processing. The data we are interested in was extracted as follows. The image was imported into the bitmap editor GIMP. A transparent layer was applied on top of it, which we worked with later. The brush size was set to match the size of the particles in the image~\cite{Park2006}. All particles not included in the annular sediment were drawn on the transparent layer manually on top of the experimental image. At the same time, we tried to choose different colors for each cluster for the convenience of further calculation.
\begin{algorithm}[H]
\caption{Algorithm for finding clusters.}
\label{alg:clustering}
    \begin{algorithmic}[1]
      \State Read the coordinates of the particles $x_i$, $y_i$ and their cluster numbers  $c_i \leftarrow 0$ ($i \in [1;N_p]$).
     \For {$i \leftarrow 1, N_p$}
            \State $r\leftarrow \sqrt{x_i^2 + y_i^2}$
            \If {$r>r_*$}
            \State remove this particle from the array.
            \EndIf
      \EndFor
       \State Cluster counter $C_n \leftarrow 1$.
      \For {$i \leftarrow 1, N_p$}
        \For {$j \leftarrow 1, N_p$}
            \State Calculate the distance $s_{ij}$ between particles $i$ and $j$.
            \If {$s_{ij}=d_p$ $\And$ $i \neq j$}
                \If{$c_i= 0$  $\And$ $c_j= 0$ }
                    \State $c_i\leftarrow c_j\leftarrow C_n$
                    \State $C_n \leftarrow C_n +1$
                \EndIf
                \If{$c_i \neq 0$ $\And$ $c_j= 0$}
                    \State $c_j \leftarrow c_i$
                \EndIf
                \If {$c_i= 0$ $\And$ $c_j \neq 0$}
                    \State $c_i \leftarrow c_j$
                \EndIf
            \EndIf
        \EndFor
      \EndFor
      \While{there are changes to cluster numbers}
      \For {$i \leftarrow 1, N_p$}
        \For {$j \leftarrow 1, N_p$}
        \State skip the zero cluster particles.
        \State $s_{ij}= \sqrt{(x_i-x_j)^2 + (y_i-y_j)^2}$
        \If {$s_{ij}=d_p$ $\And$ $i \neq j$}
        \If {particles $i$ and $j$ are not from one cluster}
            \State We assign a value $\max(c_i,c_j)$ to the particle with $\min(c_i,c_j)$.
            \EndIf
        \EndIf
         \EndFor
      \EndFor
      \EndWhile
    \end{algorithmic}
\end{algorithm}

Sometimes it was difficult to visually determine whether a particular particle belongs to a particular cluster or not. In the process, we were guided by the following rule. If a particle is located close to another particle in the cluster, or the distance between these particles is much smaller than the particle size, then we assume that the current particle also belongs to this cluster. As a result, the layer with the original image was removed. Based on the received image, we manually calculated the average cluster size.

Processing of numerical simulation results was performed using the algorithm~\ref{alg:clustering} for finding clusters.
To determine the contact of two particles, the condition $s = d_p\pm \varepsilon d_p$ was used, where $s$ is the distance between two particles, the diameter of a spherical particle $d_p=2r_p$, and $\varepsilon$ is the constant ($\varepsilon=$ 0.01 is used here). This nuance was also taken into account when detecting a collision of two particles ($s < d_p- \varepsilon d_p$). The calculation of the average cluster size based on the results of ten repeated tests was performed using a script written in Python.

\section{Results and discussions}
\begin{figure}[h]
\begin{minipage}[c]{0.7\textwidth}
  \centering
  \includegraphics[width=0.99\textwidth]{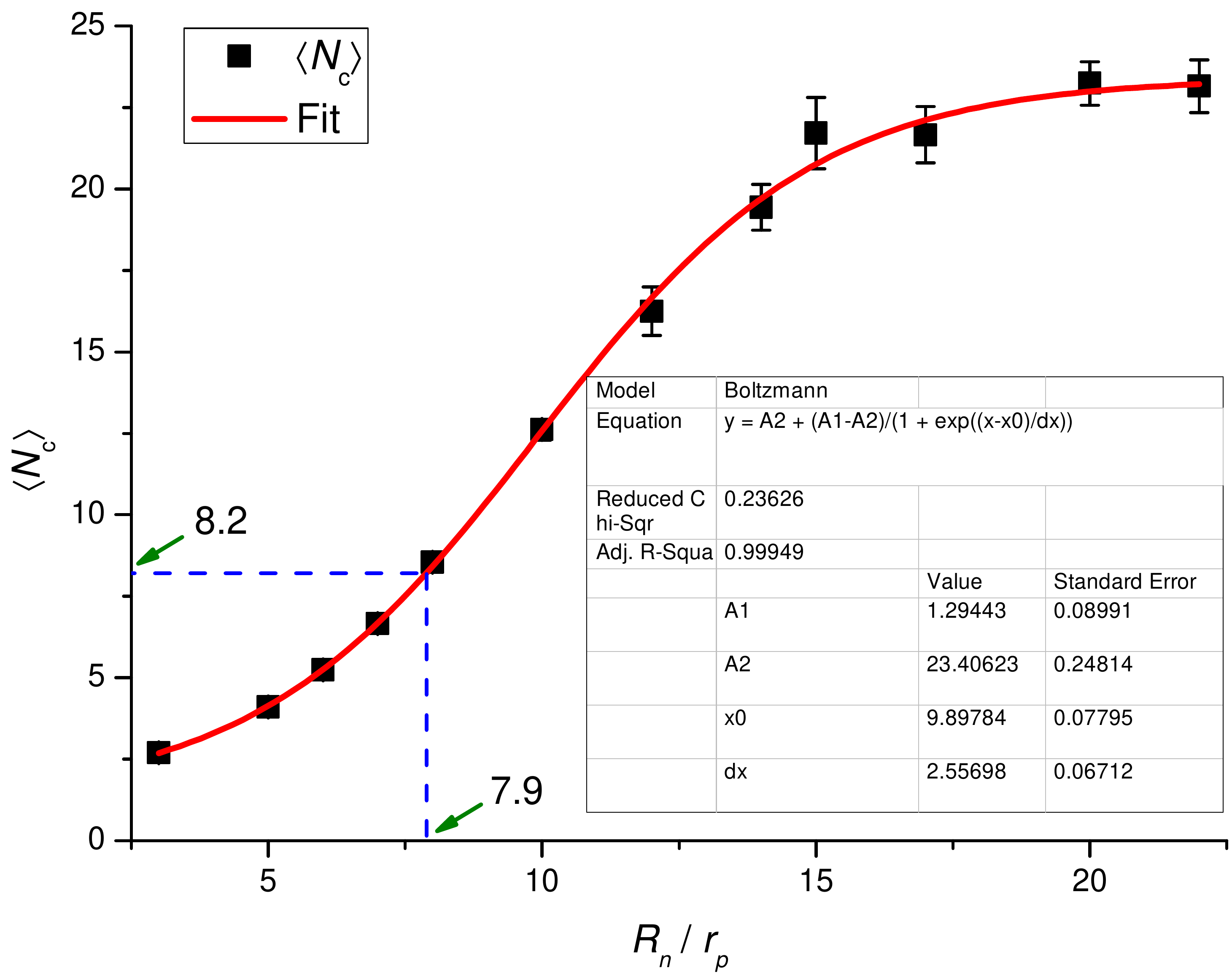}\\
\end{minipage}
\begin{minipage}[c]{0.3\textwidth}
\caption{\label{fig:averageNc_vs_Rn}Dependence of the average cluster size on the parameter $R_n$ (the standard deviation was used to estimate the error).}
\end{minipage}
\end{figure}

A series of calculations were performed for different values of $R_n$ in the range from $3r_p$ to $22r_p$. Then  the numerical result was approximated using the least-squares method (Figure~\ref{fig:averageNc_vs_Rn}). The plot in Figure~\ref{fig:averageNc_vs_Rn} resembles a sigmoid. The value of the average cluster size determined by us based on experimental data~\cite{Park2006} is $\langle N_c \rangle \approx 8.2$ (average number of particles in the chain). The model predicts this value for $R_n \approx 7.9 r_p$.

\begin{figure}[h]
\begin{minipage}[c]{0.35\textwidth}
\includegraphics[width=0.99\textwidth]{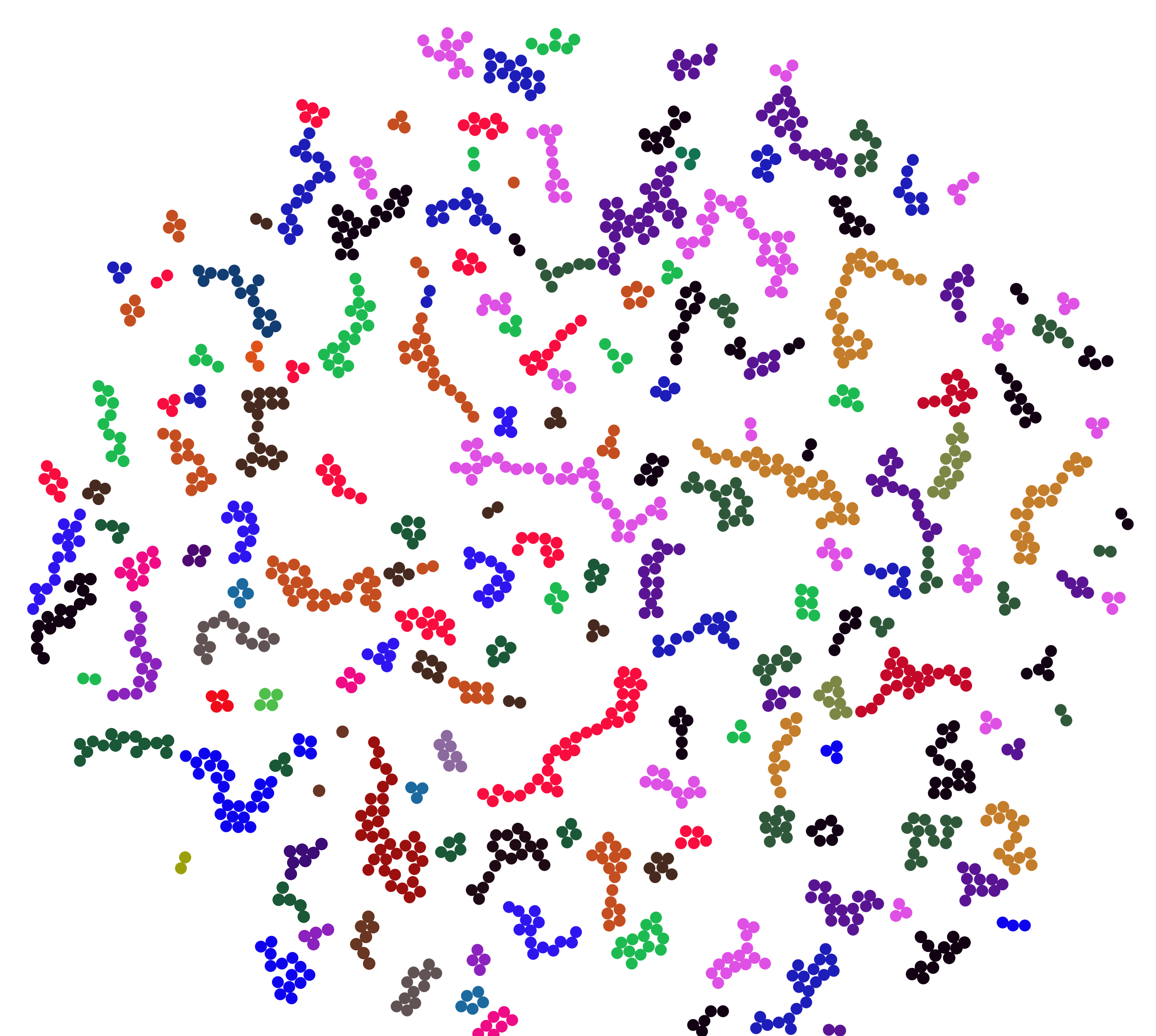}
\end{minipage}
\begin{minipage}[c]{0.35\textwidth}
\includegraphics[width=0.99\textwidth]{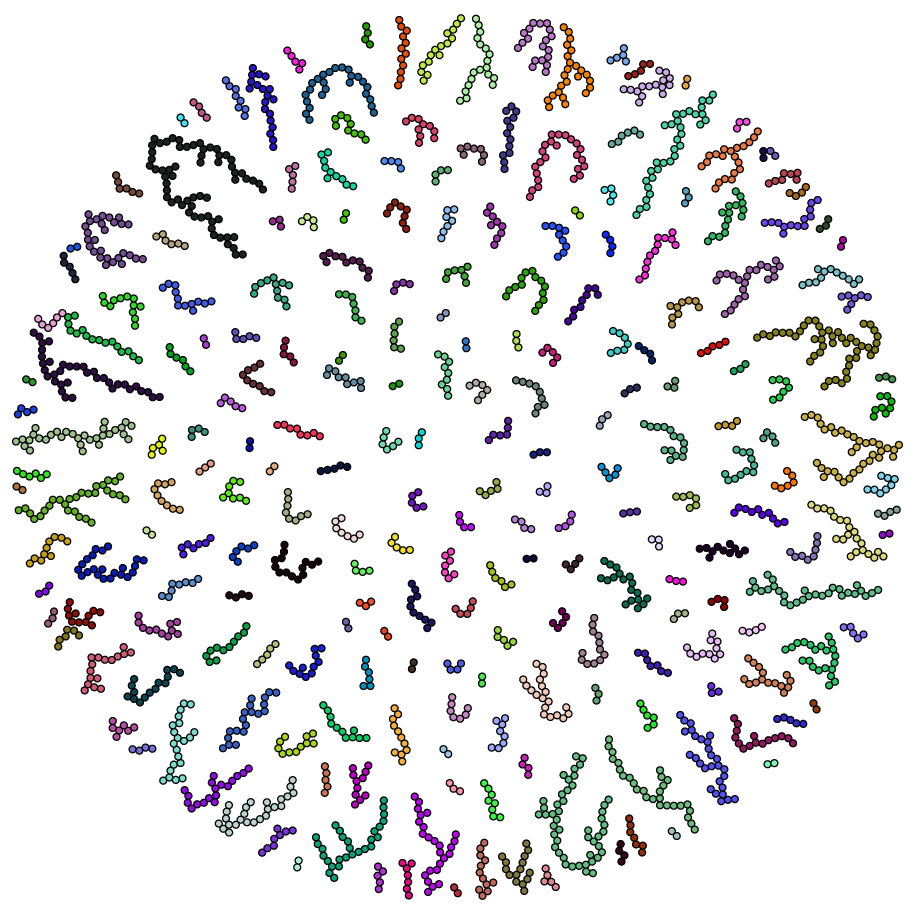}
\end{minipage}
\begin{minipage}[c]{0.25\textwidth}
	\caption{\label{fig:experimentAndSimulationClusters}  Visualisation of clusters from experiment~\cite{Park2006} (left) and simulation results for $R_n = 8 r_p$ (right).}
\end{minipage}
\end{figure}

For comparison, the Figure~\ref{fig:experimentAndSimulationClusters} shows the final sediment structures based on experimental data~\cite{Park2006} and our simulation results. Only the particles located inside the annular sediment are shown. The subset of particles belonging to the ring was subtracted from their set using the condition $r>r_*$. The inner radius of the ring $r_*\approx 0.75R$ was determined based on calculation data~\cite{Kolegov2019} ($r_*$ corresponds to the place of a high gradient of the number of particles per unit area).   The internal structure of such rings was discussed in detail earlier~\cite{Kolegov2019}.

Here we observe that for the given parameters, the model predicts the absence of single particles (Figure~\ref{fig:experimentAndSimulationClusters}). The visualization of experimental data shows a very small number of such particles. In both cases, there are clusters of different sizes (from two to several dozen particles in a cluster). Some clusters are similar in shape. It should be noted that based on the results of modeling, tree-shaped clusters are expected to appear. Their root element is located closer to the periphery. These ``trees'' grow towards the center of the area. It is explained by the nature of movement and the shape of the fixation boundary $R_f$. Such branching structures are not observed in the experiment. This distinctive feature of the numerical results is most likely related to the approximate description of the domain geometry~\cite{Kolegov2019}. After all, the shape of the drop at the final stage of evaporation is more like a film, in which local ruptures can occur~\cite{Kolegov2020}. Thus, the action of capillary forces occurs in a larger area. The model boundary $R_f$ and its neighborhood are only a rough approximation of such a region. Besides, the mimic of the capillary interaction of particles is used here. In a good way, it is necessary to solve the n-body problem numerically. Only in this problem, instead of gravity, there are capillary forces.

\begin{figure}[h]
	\begin{minipage}[c]{0.7\textwidth}
		\centering
		\includegraphics[width=0.99\textwidth]{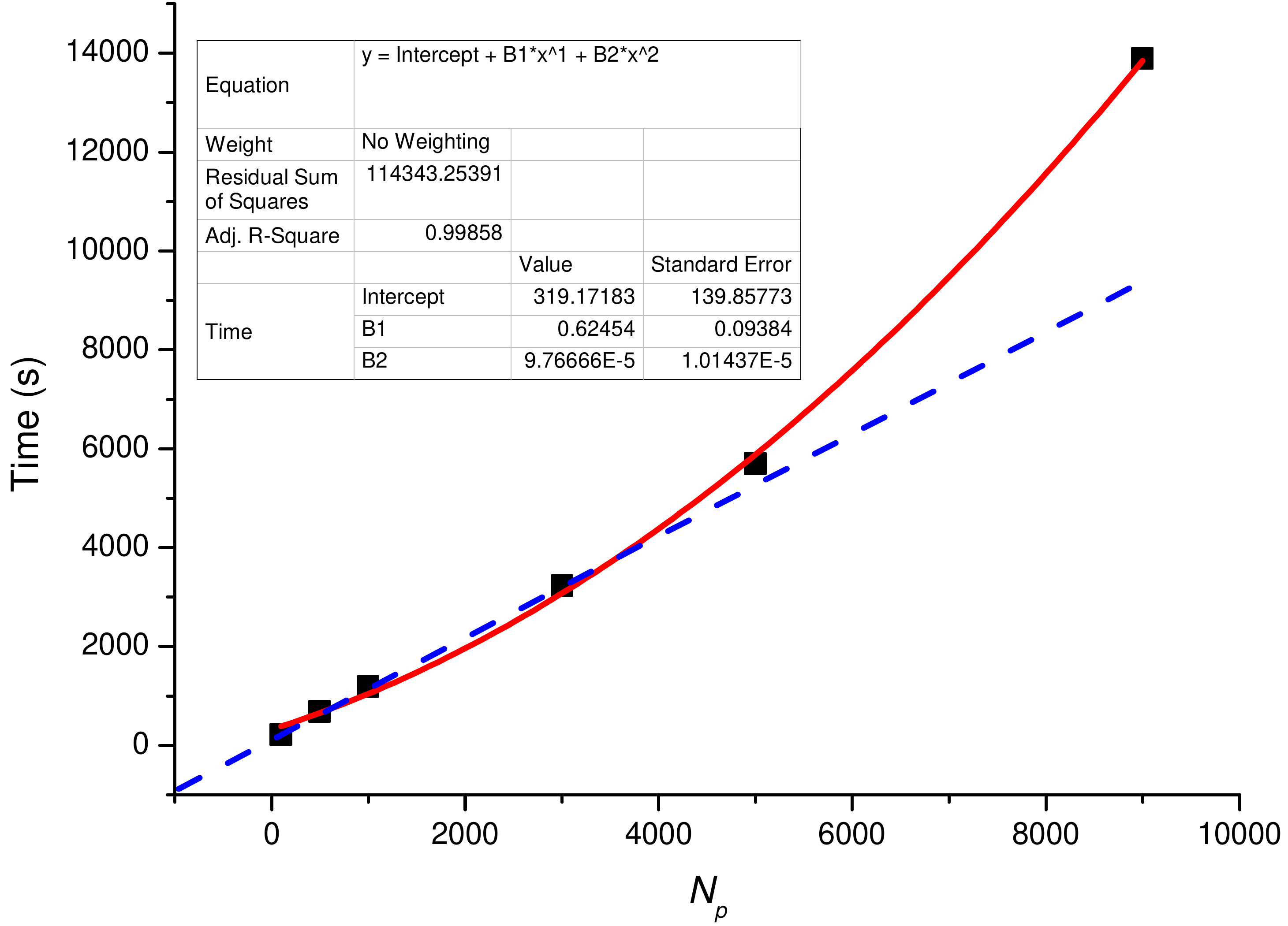}\\
	\end{minipage}
	\begin{minipage}[c]{0.3\textwidth}
		\caption{\label{fig:NpVsTime}Dependence of the simulation time on the particle number (the program was written in Python, Intel (R) Core (TM) i9-9900K CPU 3.6 GHz was used).}
	\end{minipage}
\end{figure}

Analysis of the algorithm~\ref{alg:particleDynamicsModified} showed a weak nonlinearity of the dependence of the simulation time on the number of particles $N_p$ (Figure~\ref{fig:NpVsTime}). Also, it was found that the average (over time steps) rate of movable particle displacement failures for the modified algorithm (with multiple attempts to particle displacement) is 1--2\% according to the results of calculations. For comparison, we tested the original algorithm (with a single attempt to shift the particles) and got a failure rate of 3-4\%. Five tests were run simultaneously for each type of calculation (the error is less than the marker size on the Figure~\ref{fig:NpVsTime}). Thus, for a sufficiently small value of the time step, the error associated with the failure of the particle shift is small. Repeated attempts to shift particles can slow down the program and not give a significant difference in the result.

\section{Conclusion}
Numerical results~\cite{Kolegov2019} are qualitatively agreed with the experiment~\cite{Park2006}. As the droplet dries, chains of particles (clusters) are formed inside the annular sediment. In the current work, we have processed experimental data~\cite{Park2006}, found out the average cluster size, and defined the value of a model parameter $R_n$  to predict a quantitatively agreed value $\langle N_c \rangle$. But it is also worth mentioning the qualitative difference between the simulation and experimental results, related to the shape of some clusters. The simple model under consideration often predicts tree-shaped clusters, which is not typical for the experiment~\cite{Park2006}. Further study is needed in this direction to better understand the processes that are taking place. Some questions can be answered by using the high-speed shooting of the formation of such clusters. Additional experimental data are also needed, for example, on the shape of the free surface of the liquid at the time stage when these clusters begin to appear. Further creation of more complex and accurate models that can predict the internal structure of such precipitation is important.

\section*{Acknowledgment}
This work is supported by the grant 18-71-10061 from the Russian Science Foundation. We thank Svetlana Kolegova for her help in processing experimental data and Yuri Tarasevich for useful comments.


\section*{References}
\bibliography{ref}

\providecommand{\newblock}{}
\begin{thebibliography}{10}
\expandafter\ifx\csname url\endcsname\relax
  \def\url#1{{\tt #1}}\fi
\expandafter\ifx\csname urlprefix\endcsname\relax\def\urlprefix{URL }\fi
\providecommand{\eprint}[2][]{\url{#2}}

\bibitem{Kolegov2020}
Kolegov K~S and Barash L~Y 2020 {\em Advances in Colloid and Interface
  Science\/} {\bf 285} 102271 ISSN 0001-8686

\bibitem{ZhaoM2019}
Zhao M, Luo W and Yong X 2019 {\em Journal of Colloid and Interface Science\/}
  {\bf 540} 602--611

\bibitem{Wouters2020}
Wouters M, Aouane O, Sega M and Harting J  (\textit{Preprint}
  \eprint{2007.15405v2})

\bibitem{Chun2020}
Chun B, Yoo T and Jung H~W 2020 {\em Soft Matter\/} {\bf 16} 523--533

\bibitem{Seo2020}
Seo H~W, Jung N and Yoo C~S 2020 {\em Journal of Mechanical Science and
  Technology\/} {\bf 34} 801--808

\bibitem{Darras2017}
Darras A, Opsomer E, Vandewalle N and Lumay G 2017 {\em Scientific Reports\/}
  {\bf 7}

\bibitem{Yang2020}
Yang J, Kim H, Lee C, Kim S, Wang J, Yoon S, Park J and Kim J 2020 {\em
  Theoretical and Computational Fluid Dynamics\/} {\bf 34} 679--692

\bibitem{Deegan1997}
Deegan R~D, Bakajin O, Dupont T~F, Huber G, Nagel S~R and Witten T~A 1997 {\em
  Nature\/} {\bf 389} 827--829

\bibitem{Kolegov2019}
Kolegov K~S and Barash L~Y 2019 {\em Physical Review E\/} {\bf 100}

\bibitem{Park2006}
Park J and Moon J 2006 {\em Langmuir\/} {\bf 22} 3506--3513

\end{thebibliography}

\end{document}